# TRANSMISSION OF VIDEO IMAGE

# IN UNDERWATER ACOUSTIC COMMUNICATION

**Han Zhang**[a,b,*] , **Jun Yang**[a,b,c]

[a] State Key Laboratory of Acoustics, Institute of Acoustics, Chinese Academy of Sciences, Beijing 100190,China

zhanghan@mail.ioa.ac.cn

[b] Key Laboratory of Noise and Vibration Research, Institute of Acoustics, Chinese Academy of Sciences, Beijing 100190,China

[c] School of Electronic, Electrical and Communication Engineering, University of Chinese Academy of Sciences, Beijing 100049, China

**Abstract**

Acoustic communication is currently considered as the best way to transmit information over long distances under water, since acoustic waves have lower attenuation in water than the other information transmission media. However, current information transmission rates of the underwater acoustic communication is usually limited by low-frequency bandwidth due to the large attenuation which characterized in high-frequency sound waves in water. Spiral acoustic beams with helical wavefront dislocations carrying orbital angular momentum have been applied to acoustic levitation and capture. Recently, harnessing the orthogonality and the infinite dimension of Hilbert space in underwater acoustic communication has been considered promising. Here, we construct a real-time underwater acoustic communication system to transmit video image. The system multiplexes eight orbital angular momentum topology charges ranging from +1 to +8 and encoded with on-off keying modulation format, achieving a spectral efficiency of 8 (bit/s)/Hz. This demonstration suggests that multiplexed information-carrying orbital angular momentum acoustic beams presents tremendous potential for increasing the capacity of underwater communication systems.

* This work is supported by the National Natural Science Foundation of China (Grant 1177021304).

**Introduction**

A particular acoustic waves carrying orbital angular momentum (OAM) is known as vortex acoustic beams, which characterized by a screw wavefront - that is, surface of the constant phase - dislocation[1,2]. In the wavefront dislocation of a screw monochromatic, continuous traveling acoustic wave beam, there is a phase dependence on the azimuthal angle $\theta$, which can be formed of $\exp(\mathrm{il}\,\theta)$[3], where the integer $1$ is called topological charge or the order of the screw beam whose sign and value determines the orientation of the spiral and the magnitude of the OAM, respectively.

Related research shows that there is a perfect orthogonality relation between vortex acoustic beams with different OAM topological charges[4], so that the multiplex vortex beams can be transmitted independently. In recent years, orthogonal acoustic vortex beams with orbital angular momentum is considered bringing a new degree of freedom having a potential to increase the information capacity substantially and improve the ability of acoustic waves for communication[4], especially combining with multilevel amplitude/phase modulation formats.

At present, the commonly used acoustic communications is the utilization of the plane acoustic waves[5-10] (far-field approximation of a spherical wave) whose wavefront is perpendicular to the propagation axis. Such acoustic waves have been utilized since the beginning of the acoustic communication several decades ago. So far, there is no much change in the acoustic waves. At the receiving terminal, the intensity of the acoustic pressure field is detected, whose velocity of change indicates the frequency and sequence of change represents the phase. Available dimensions have been limited to the domain of time, frequency and space, and if there is no new degree of freedom can be utilized, a significant increase of channel capacity will be difficult, according to the Shannon formula in information theory.

Studies indicated that OAM is another physical quantity that different from the intensity of acoustic pressure field[11], if we can take advantage of acoustic OAM, the acoustic wave transmission and detection capabilities could be greatly enhanced.

Vortex field carrying angular momentum is a very common phenomenon in nature, many studies have confirmed that both the electromagnetic[12-14] or the acoustic waves[15,16] can form a vortex field under certain conditions. Wave front dislocation is the intrinsic property of the vortex field, which was first discussed in acoustics[2]. However afterward, a majority of the subsequent researches are conducted in optics[17-21] for which is supported by a solid foundation of the quantum mechanics and the fundamentally available circular polarization of spin which is not exist in acoustic. Allen et al[12]. first proved that the optical vortex has a well-defined OAM when propagated along the paraxial axis according to the Maxwell's equations. This conclusion has drawn a great deal of attention from scientists. From then on, optical OAM applied in communication has been made significant progress. Inspired by optical OAM, acoustic OAM has revived, attracting scientists' research once again in recent years.

**Principle**

In the field of optics and microwaves, helical beams with different OAM topological charges can be generated by spatial light modulators (SLM), metamaterials, or parity-time symmetric ring resonators and multiplexed by beam splitters or spin-orbit coupling, significantly improve the communication capacity[18,22-27].

In optics, a particular solution satisfies the paraxial wave equation that can describe for the complex scalar function of a optical vortex mathematically is known as Laguerre-Gaussian (LG) beam[12,21,28] which could be conveniently express in the form of cylindrical polar coordinates $(\rho, \theta, z)$. The variation of the amplitude is

$$
\begin{aligned}
u_{1,p}(\rho, \theta, z) = & \sqrt{\frac{2p!}{\pi(p+|1|)!}} \frac{1}{w(z)} \left(\frac{\rho\sqrt{2}}{w(z)}\right)^{|1|} \exp\left(\frac{-\rho^2}{w^2(z)}\right) L_p^{|1|}\left(\frac{2\rho^2}{w^2(z)}\right) e^{i1\theta} \\
& \times \exp\left(-ik\frac{\rho^2 z}{2(z_R^2+z^2)}\right) \exp(-i\phi)
\end{aligned}
\tag{1}
$$

where natural number $p$ is the radial index affecting the number of radial nodes of the wave, $L_p^{|l|}$ is the associated Laguerre polynomial, $k = 2\pi/\lambda$ is the wavenumber, $z_R = kw^2(0)/2$ is the Rayleigh range indicating the tightness of the focus, $w(z) = w(0)\sqrt{1 + z^2/z_R^2}$ is the local beam width, $w(0)$ is the beam waist and $\varphi = (2p + |l| + 1)\tan^{-1}\left(\dfrac{z}{z_R}\right)$ is the Guoy phase shift. From the term of $e^{il\theta}$, one can observe a azimuthal phase dependence about the z-axis.

Inspired by optics, we can introduce a analogous LG beam expression of acoustic vortex field[29,30], for the present analysis, considering with $p = 0$, and neglect the transmission loss i.e. $u(\rho,\theta,z) = u(\rho,\theta,0)$. Thus we can obtain a complex wave function $\psi$ of the acoustic vortex with a quasi-LG expression

$$
\begin{aligned}
\psi(\rho,\theta,z;1,\omega,t) &= u_{1,0}(\rho,\theta,0)e^{i(\omega t + kz + 1\theta)} \\
&= \sqrt{\frac{2}{\pi|1|!}}\frac{1}{w(0)}\left(\frac{\rho\sqrt{2}}{w(0)}\right)^{|1|}\exp\left(\frac{-\rho^2}{w^2(0)}\right)L_0^{|1|}\left(\frac{2\rho^2}{w^2(0)}\right)e^{i(\omega t + kz + 1\theta)}
\end{aligned}
\tag{2}
$$

So that the intensity distribution of the quasi-LG beam is obvious

$$
\begin{aligned}
I_{1,0} &= \left|u_{1,0}(\rho,\theta,0)\right|^2 \\
&= \frac{2}{\pi|1|!w^2(0)}\left(\frac{2\rho^2}{w^2(0)}\right)^{|1|}\left[L_0^{|1|}\left(\frac{2\rho^2}{w^2(0)}\right)\right]^2\exp\left(-\frac{2\rho^2}{w^2(0)}\right)
\end{aligned}
\tag{3}
$$

then we can find the location of the maximum intensity shaping a circular profile by solving $I_{1,0} = (I_{1,0})_{max}$, i.e. the radius $R(R > 0)$ of the circularity satisfies the equation

$$\left(\frac{2}{|1|}\right)^{|1|}\left(\frac{R}{w(0)}\right)^{2|1|}e^{-\frac{2R^2}{w(0)^2}}=e^{-|1|} \qquad (4)$$

according to which, radius of the maximum intensity position with different 1 can be obtained.

The emission field from an acoustic source can be expressed as a Gaussian beam model[31], hence we can approximately simulate a quasi-LG acoustic vortex beam by discretizing the circularity where the maximum intensity located with desired topological charge and arrange into $n$ acoustic OAM array elements. Specifying a reference array element and rotating in a fixed direction (clockwise or counterclockwise), the applied incentive signal obeys that

$$U_i(t)=I_{1,0}^{\max}\exp\left[i\left(\omega t+1\,\theta_i\right)\right] \qquad (5)$$

$$\theta_i=\frac{2\pi}{n}i \qquad (6)$$

where $i=0,1,2,...,n-1$ denotes the number of the acoustic array element, $\theta_i$ is the azimuthal angle between the number $i$ array element and reference acoustic source.

The OAM beams carrying information can be described as

$$U_i^S(t)=S(t)\cdot U_i(t)=S(t)I_{1,0}^{\max}\exp\left[i\left(\omega t+1\,\theta_i\right)\right] \qquad (7)$$

where $S(t)$ is the baseband digital signal to be sent.

$\ell=+1$    $\ell=+2$    $\ell=+3$    $\ell=+4$

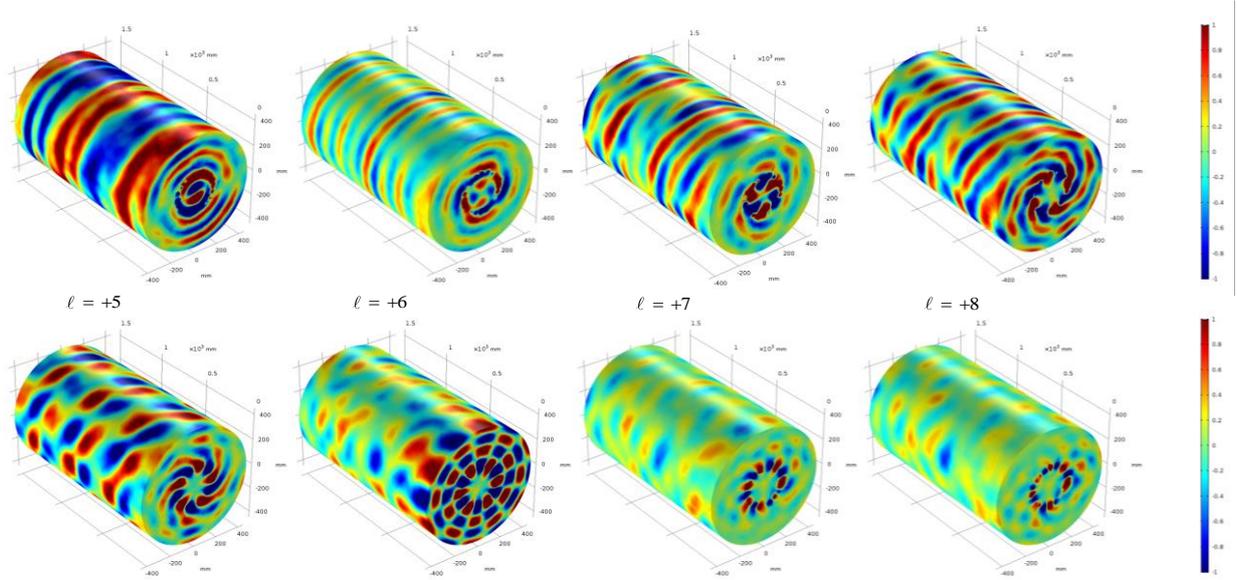

**Figure 1 | OAM spiral acoustic fields with topological charge from +1 to +8 driven by a 20-elements circular array. Culour bar represents normalized pressure.**

## Multiplexed OAM beams

Different from optics[32-34], acoustic wave has no polarization characteristics, leading to non-rotation state phenomenon in acoustic field. However, after introducing the concept of pseudo angular momentum[35,36], a connection between the optical vortices and acoustic vortices can be established. It is demonstrated that the topological charge of a wave vortex is associated with the conservation of pseudo angular momentum in the linear regime resulting from the medium isotropy, while for the weak nonlinear regime, pseudo angular momentum is conserved when the ratio between the topological charge and the frequency remains constant, which is of great potential significant for acoustic communication.

The introduction of the spatial angular momentum degree of freedom enables us to increase the channel capacity of acoustic communications significantly through the multiplexing of acoustic OAMs beams with different topological charges. In optics, the vortex beams with certain OAM topological charge is generally generated by means of passive measures such as spatial light modulator[22] and micro-resonator ring[18], and multiplexed through beam splitter or spin-orbital coupling. While for acoustic, passive modulation structures[29,37-39] can only product single topological charge vortex beams and hard to perform the multiplex operation via any media structures, and acoustic beam splitter is still a challenging research topic. Hence, acoustic vortex beam is usually emitted by active phase arrays[1,15,40], and pre-multiplex by means of digital signal processing in digital domain to obtain a series of excitation signals driving the corresponding array elements.

For the multiplexing of $N$ OAM channels carrying information, the multiplexed excitation signal of the number $i$ element of the emitting plane is

$$U_i^{MUX}(t) = \sum_{l=1}^{N} S_l(t) I_{l,0}^{\max} \exp\left[ i\left( \omega t + l \theta_i \right) \right] \qquad （8）$$

where OAM topological charges $l = \pm 1, \pm 2, \pm 3, ..., \pm N$. Although the OAM acoustic vortex beams overlap with one another when transmitting multiplexed, information carried by each of the OAM acoustic beams remains independence with each other because of the mutual orthogonality between vortex beams with different OAM charges.

**Demultiplexing OAM beams**

Demultiplexing is still in the digital domain. A down-sampling array of a suitable resolution is utilized for acoustic field reception, recording both amplitude and phase information of the received acoustic field, and then post-process the received data through digital signal processing.

According to the intrinsic orthogonality in the OAM beams[4], one can demultiplex the received information into original several independent channels conveniently by multiplying a inverse spiral phase term $\exp\left[i\left(-1_q\theta\right)\right]$

$$
\begin{aligned}
U^{DEMUX}(t) &= \exp\left[i\left(-1_q\theta\right)\right]\cdot\sum_{p=1}^{N}S_p(t)\cdot I_{1_p,0}^{\max}\exp\left[i\left(\omega t+1_p\theta\right)\right] \\
&= S_q(t)\cdot I_{1_q,0}^{\max} + \sum_{p=1,p\neq q}^{N}S_p(t)\cdot I_{1_p,0}^{\max}\exp\left[i\left(\omega t+1'_p\theta\right)\right]
\end{aligned}
\tag{9}
$$

where $1'_q = 1_p - 1_q$. After that, the OAM topological charge $1_q$ back-converts into the plane acoustic beam with no wavefront dislocation, i.e., $1_q$ degenerates into zero distinguishing from the other OAM beams which are characterized by null core. So the information attached to the $1_q$ can be extracted after filtering out the second term.

**Experimental setup**

We conducted an underwater static acoustic communication scheme (Fig. 2) applying OAM multiplexing combining with a simple modulation format on-off keying (OOK) via numerical simulations.

The active acoustic array is set on a single ring distributed uniformly with 20

transducers is used to generate and modulate the OAM acoustic beams applying the phased array technology and emit multiplex acoustic beam with OAM topological charges +1 to +8 into free space at 10 kHz, symbol period 4 times the carrier period (i.e. baud rate is 2.5 kHz). The radius of the ring is set to one wavelength and the received acoustic pressure field is measured at 20 wavelengths away from the emitting plane in the axial direction.

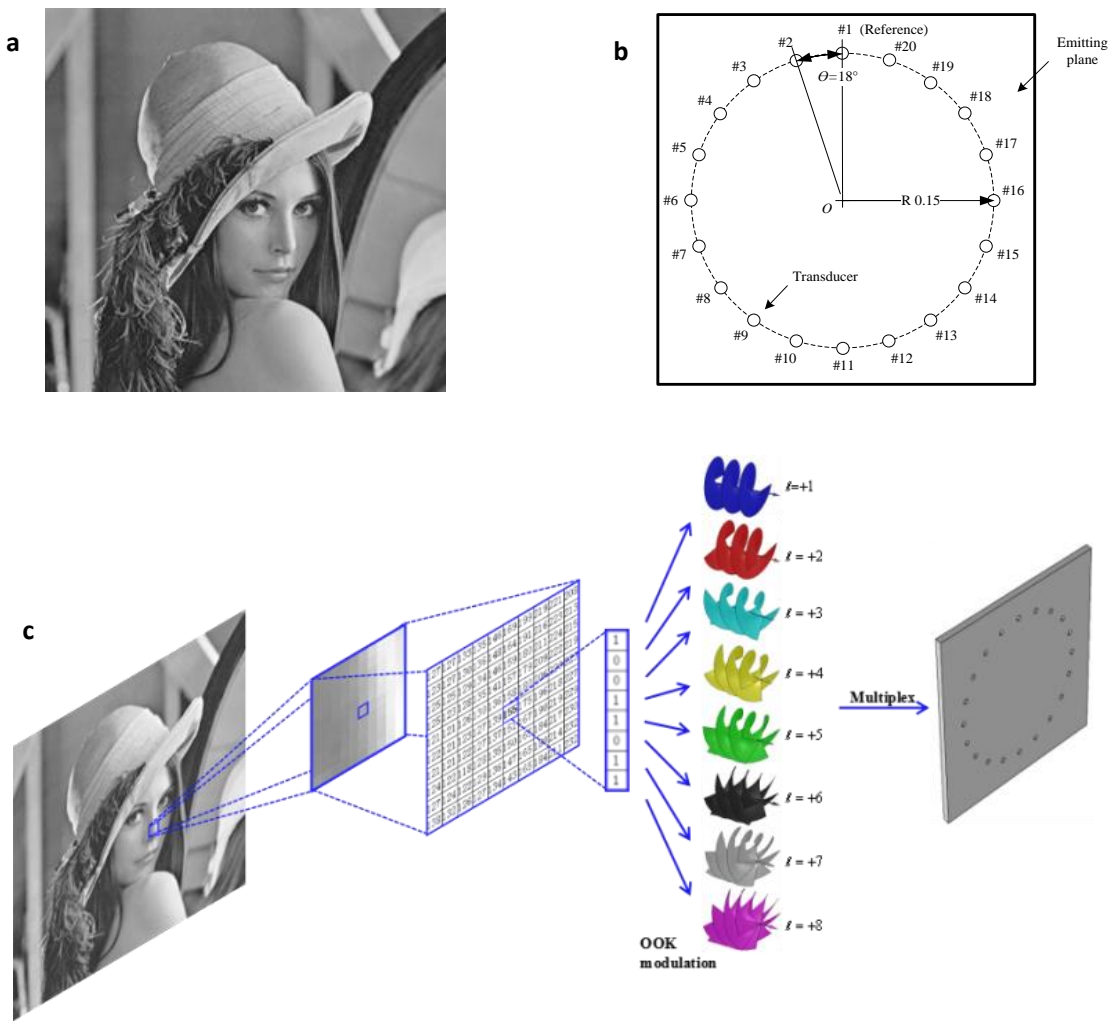

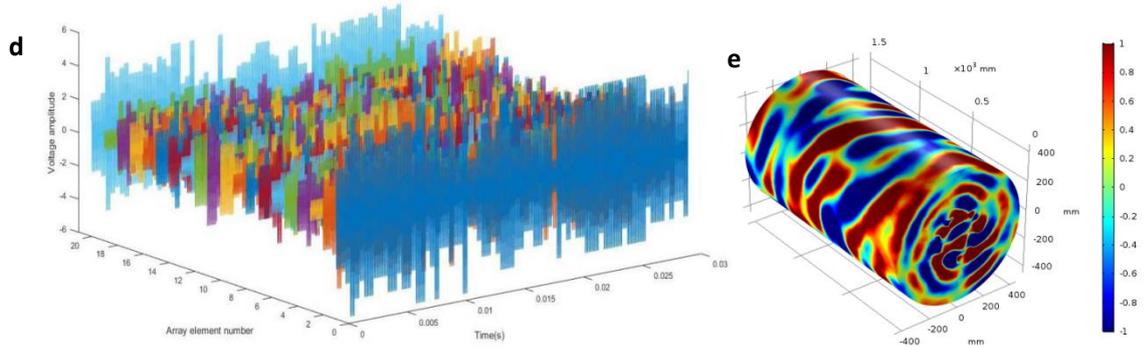

**Figure 2 | Notion of underwater acoustic communication system multiplexing 8 OAM topological charges. a,** The gray scale image with 256×256 pixels of lena (lena.jpg) to be transmitted through the communication system. **b,** The emitting plane composed of 20 transducers assigned in an circle with the radius of one wavelength. **c,** An image is composed of pixels, which is presented by an 8-bit binary number. using multiplexed acoustic OAM beams, here the number of channel is 8, and under the modulation of OOK, information can be transmitted by parallel. **d,** Driving signals in the 20 transducers for generating the multiplexing channels of the eight type of OAM. The signals are in amplitude modulation and only presented for a little part in time. **e,** Simulation of the pressure field when transmitting the pixel of "155" (i.e. "10011011"in binary format c.f. Fig 2b) at 6 ms. Colour bar represents normalized pressure.

In the simulation, we implemented the transmission of a standard gray scale image of Lena, the size of which is 256×256 pixels. For each pixel is encoded as 8-bit binary data, we can independently encode the pixels of the image into the OAM acoustic carrier wave in OOK format with each bit occupying an independent specified OAM channel. The image data is transmitted in parallel through multiplexed OAM beam.

At the receiving terminal, the acoustic pressure field is collected through a 8*8

receiver array in a scan area of 1.5 wavelengths by 1.5 wavelengths (Fig. 3a), and the recorded data of both amplitude and phase is then performed products with the spiral phase reversed of the eight bases to demultiplex and restore the original baseband signal.

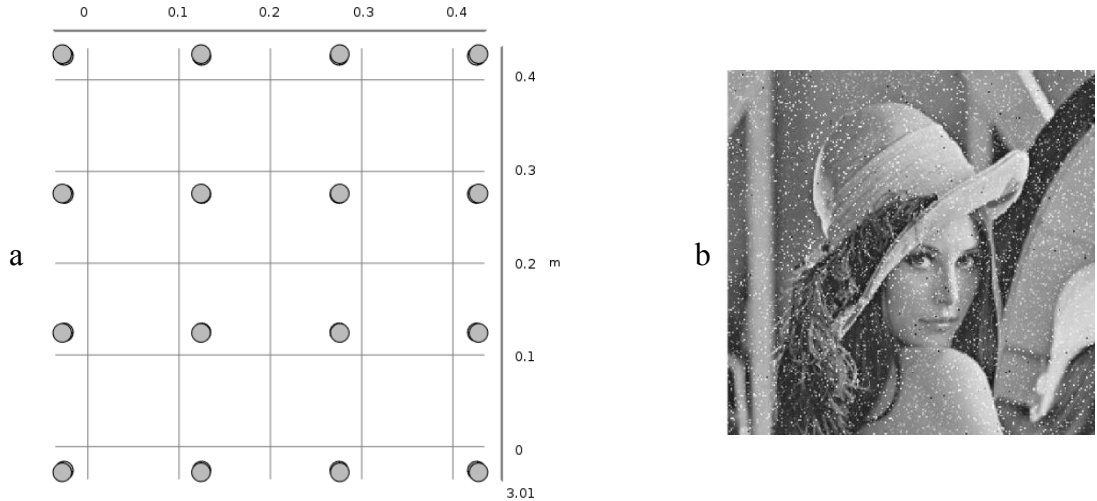

**Figure 3 | The receiving terminal in the experimental setup. a,** The receiver array recording the amplitude and phase of acoustic pressure field. **b,** The receiving image obtained by forming the demultiplex algorithm at 20 dB signal-to-noise ratio (SNR) where additive white gaussian noise (AWGN) was only concerned.

**Discussion**

The experimental setup presented here is very versatile and the acoustic vortex spatial and temporal patterns are configurable in real time.

We studied the effect of the scanning area and receiver resolution on the performance of the communication system (Fig 4). Three experimental setups of 8*8 full area, 8*4 half area and 16*4 half area were performed. And the result shows that the original information can also be demodulated from the received acoustic field of multiplexed

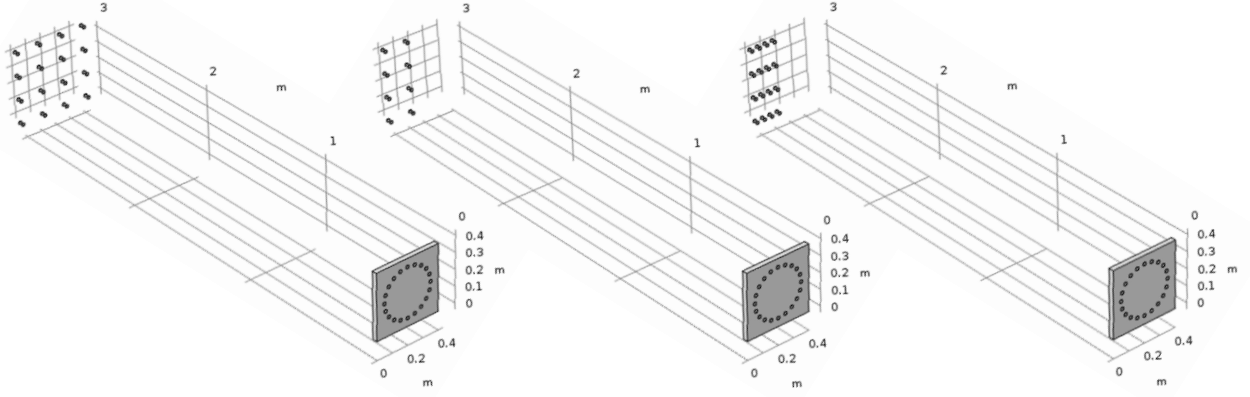

OAM beams in half area. However, even if the resolution is increased to keep the same number of sensors as the original full area, the bit error rate also increases, and the effect becomes worse when the overall effect is more comprehensive.

a                                b                                c

**Figure 4 | Three receiving configurations. a,** 4×4 receiver array set in the scan area of 1.5 wavelengths by 1.5 wavelengths, **b,** 4×2 receiver array set in half area and **c,** 4×4 receiver array set in half area 20 wavelengths away from the emitting plane.

In addition, the relationship between the number of elements and the number of available OAM channels is also studied. The formula is derived first. Since the practical system is the discretization of the circumference, leading to a restriction in the number of OAMs that can be generated. Because of the applied incentive signal has the form of $U_i(t) = I_{1,0}^{\max} \exp\left[\mathrm{i}\left(\omega t + 1\,\theta_i\right)\right]$, so that

$$0 < 1\,\theta_i \leq 2\pi \qquad\qquad （10）$$

for $\theta_i = \dfrac{2\pi}{n}i$, we have

$$0 < 1 \leq n \qquad\qquad (11)$$

obtaining the constraint of the available OAMs that a practical system can generate.

When the number of elements is constant, the OAM channels that can be generated is determined. The theoretical limit of available OAMs is the number of the elements, but in fact the limit is smaller, reaching a certain number of multiplex channels will result in the dramatic deterioration of receiving terminal when performing the demultiplexing operation, and can not restore the original signal correctly.

The theoretical limit of the spectrum efficiency given by the number of OAM topological charges $N$ used for data transmission, which in this experiment is 8 (bit/s)/Hz. And the information transmission rate is increased by 8 times compared with the direct OOK modulation. Like other multiplexing methods, OAM multiplexing can also be combined with existing cutting-edge advanced modulation formats to increase $N$ times the communication rate of underwater acoustic communication systems. On the other hand, theoretical communication capacity expansion is unlimited due to the characteristic of infinite dimensional Hilbert space in OAM. Even when limited by practical physical constraints, the acoustic communication rate can also be substantially improved according to the number of utilized OAM topological charges, making it the most promising novelty technology that can break the bottleneck of the current underwater acoustic communication rate.

**Conclusion**

An active acoustic array is simulated to generate quasi-Laguerre Gaussian spiral acoustic beams and the driving signals of each transducers at the circular array are derived. Utilizing OAM topology charges from +1 to +8, we design an underwater acoustic OAM multiplexing communication system with modulation format OOK to send a grayscale image of Lena, making the spectral efficiency reach up to 8 (bit/s)/Hz, which can be further improved by incorporating more topological charges in combination with advanced modulation formats. Multiplexed OAM beams presents tremendous potential for improving the capacity of acoustic communication systems.

The number of available OAM channels and the receiving performance of the system are discussed. It is found that the number of usable OAM channels is related to the number of emitting transducers on the circumference, and the receiving performance is related to the number of receivers and the scanning area, regardless of the way of arrangement. The higher the resolution and the more comprehensive the coverage area, the better receiving performance.